\documentclass[conference]{IEEEtran}

\IEEEoverridecommandlockouts

\usepackage{amsmath}
\usepackage{amsfonts}
\usepackage{stfloats}
\usepackage{amssymb}
\usepackage{amsmath,amssymb,amsfonts}
\usepackage{algorithmic}
\usepackage{graphicx}
\usepackage{textcomp}
\usepackage{xcolor}
\usepackage{lettrine}
\usepackage{comment}
\usepackage{scalerel}
\usepackage{tikz}
\usetikzlibrary{svg.path}
\usepackage{algorithm2e}
\RestyleAlgo{ruled}
\usepackage{array}
\usepackage{balance}

\usepackage[hidelinks]{hyperref}
\usepackage[capitalise]{cleveref}
\usepackage{booktabs}
\usepackage{subcaption}
\usepackage{multirow}
\newcolumntype{P}[1]{>{\raggedright\arraybackslash}p{#1}} 
\newcolumntype{B}[1]{>{\centering\arraybackslash}p{#1}}
\usepackage{siunitx}
\DeclareSIUnit{\perunit}{p.u.}

% ...
\usepackage{fancyhdr}
\fancypagestyle{firstpage}{
\fancyhf{}
\fancyfoot[L]{\fontsize{8}{10}\selectfont979-8-3503-9042-1/24/\$31.00 ©2025 IEEE}

}

\begin{document}

\title{Unified Graph-Theoretic Modeling of \\ Multi-Energy Flows in Distribution Systems
\\

{\footnotesize}
\thanks{This paper is part of the Integrated Network Planning (iNeP) research project within the Northern German Living Lab (\textit{Norddeutsches Reallabor}) overarching project, funded by the German Federal Ministry of Economic Affairs and Energy (BMWE) under grant agreement no. 03EWR007O2}
}

\author{
    \IEEEauthorblockN{Marwan Mostafa, Daniel Wenser, Payam Teimourzadeh Baboli, and Christian Becker}%
    \IEEEauthorblockA{{Institute of  Electrical Power and Energy Technology,}
    Hamburg University of Technology, Hamburg, Germany}
    \IEEEauthorblockA{\{marwan.mostafa, daniel.wenser, payam.baboli, c.becker\}@tuhh.de}
}

\maketitle
\thispagestyle{firstpage}

\begin{abstract}
The increasing complexity of energy systems due to sector coupling and decarbonization calls for unified modeling frameworks that capture the physical and structural interactions between electricity, gas, and heat networks. This paper presents a graph-based modeling approach for multi-energy systems (MES), where each domain is represented as a layer in a multi-layer graph, and coupling technologies are modeled as inter-layer edges via a dedicated coupling layer. A steady-state solver based on a block-structured Newton–Raphson (NR) method is developed to jointly compute flows and state variables across all carriers. The proposed model is tested and validated on a realistic case study based on data from a German distribution network. The results demonstrate convergence, numerical accuracy, and consistent domain interaction, and demonstrate the method's applicability for system-wide analysis and its potential as a foundation for future optimizations in integrated energy systems.
\end{abstract}



\section{Introduction}
\label{Introduction}

The accelerating transition toward decarbonized, resilient, and efficient energy systems is driving the adoption of multi-energy systems (MES), where multiple energy carrier networks, \textit{e.g.}, electricity, gas, and district heating, are planned, operated, and optimized as an integrated system. Distribution-level MES, in particular, presents significant opportunities for local sector-coupling, enhancing flexibility, and decarbonization of heat and transport systems. However, the integrated operation and planning approach also brings complex interdependencies between different physical domains, necessitating new modeling paradigms capable of capturing their structural and operational interactions.

Traditionally, energy networks have been modeled and analyzed independently, using dedicated techniques such as AC or DC power flow for \textit{electricity}, the Weymouth or Panhandle equations for \textit{gas networks}, and thermal hydraulic models for \textit{district heating}. While accurate within their domains, these methods fail to capture the bidirectional influences between carriers in MES. For instance, combined heat and power (CHP) units, power-to-gas (P2G) plants, and electric boilers serve as multi-carrier coupling points that link the operation of one energy system to another.

Foundational studies have established the necessity of integrated analysis and operation of MES. Shabanpour-Haghighi and Seifi \cite{shabanpour-haghighi_integrated_2016} presented one of the earliest unified steady-state operational models for power, gas, and heating networks. Liu and Mancarella \cite{liu_modelling_2016-2} extended this perspective with detailed Sankey-based modeling of district energy systems. These efforts were further supported by Yang et al. \cite{yang_probabilistic_2018} and Shi et al. \cite{shi_generalized_2017-1}, who proposed probabilistic and steady-state models capturing multi-interactions under uncertainty and variation.

Parallel to these efforts, Newton–Raphson (NR) methods have been adapted to solve the coupled nonlinear equations that arise in multi-carrier systems. Men et al. \cite{men_coupling_2017-1} demonstrated the application of NR to gas-electric hybrid systems, while Zheng et al. \cite{zheng_variant_2022-1} proposed a third-order convergence variant specifically tuned for multi-energy flow. These methods improve convergence and numerical stability in solving coupled energy balance equations.

More recently, several studies have focused on extending traditional formulations to include novel computational frameworks. Tian et al. \cite{tian_novel_2022} developed a general power flow model for integrated systems using a modular structure. Massrur et al. \cite{massrur_fast_2018} proposed a decomposed solution framework to accelerate convergence in large-scale coupled systems. Dancker and Wolter \cite{dancker_joined_2022} employed a quasi-steady-state approach combining temporal averaging and power flow for integrated simulation.

However, these approaches often remain tightly bound to matrix-based formulations, limiting their flexibility and scalability. As MES networks grow in complexity, modular and topologically-aware structures are increasingly necessary. Graph-based modeling provides an abstraction by representing energy carriers as layered subgraphs, with inter-layer edges capturing coupling elements. Kalantar and Mazidi \cite{kalantar_graph-theoretic_2025} proposed a graph-theoretic framework to quantify resilience in microgrid-integrated distribution systems, while Kermani et al. \cite{agha_mohammad_ali_kermani_enhancing_2025} applied network analysis and simulation to enhance resilience in gas distribution systems. Markensteijn et al. \cite{markensteijn_graph-based_2020} further advanced this concept, introducing a unified graph-based structure for multi-energy flow problems. Together, these studies underscore the growing potential of graph theory as a structural foundation for MES modeling and optimization.

Despite significant progress in MES modeling and optimization, a clear gap remains in the structural unification of multi-carrier systems. Most existing approaches rely on domain-specific, matrix-based formulations that are difficult to generalize across heterogeneous networks. While some studies address energy carrier coupling, few integrate system topology directly into the formulation and solution of steady-state flows.

To the best of the authors' knowledge, no existing framework simultaneously integrates the structural and physical dynamics of electricity, gas, and heat networks within a unified multi-layer graph, nor leveraged this topology to develop a coupled NR solver. Recent planning-oriented studies highlight the importance of sector-coupling and uncertainty-aware optimization, but lack a topology-aware physical model as a computational foundation \cite{mostafa_cross-sectoral_2025}. This paper addresses this gap with three main contributions:
\begin{itemize}
    \item A unified multi-layer graph of \textit{electricity}, \textit{gas}, and \textit{heat} networks, including inter-carrier coupling devices.
    \item A comprehensive formulation of steady-state algebraic flow equations for each energy carrier within a coupled NR solver, with the Jacobian matrix derived directly from the multi-layer graph topology.
    \item Validation of solver efficiency and accuracy via benchmarking against established domain-specific tools.
\end{itemize}

Section~\ref{Model} introduces the proposed multi-layer graph modeling framework for multi-energy networks (MEN) and formulates the steady-state power flow equations. Section~\ref{Solution} describes the coupled NR solution strategy. A real-world case study is presented in Section~\ref{sec:case}, and results are discussed in Section~\ref{Results}. Finally, Section~\ref{Conclusion} summarizes the findings and outlines future works.

\section{Proposed Modeling Framework}
\label{Model}

\subsection{Graph-Theoretic Modeling of Multi-Energy Systems}

\begin{figure}[t]
    \centering
    \includegraphics[width=0.85\linewidth]{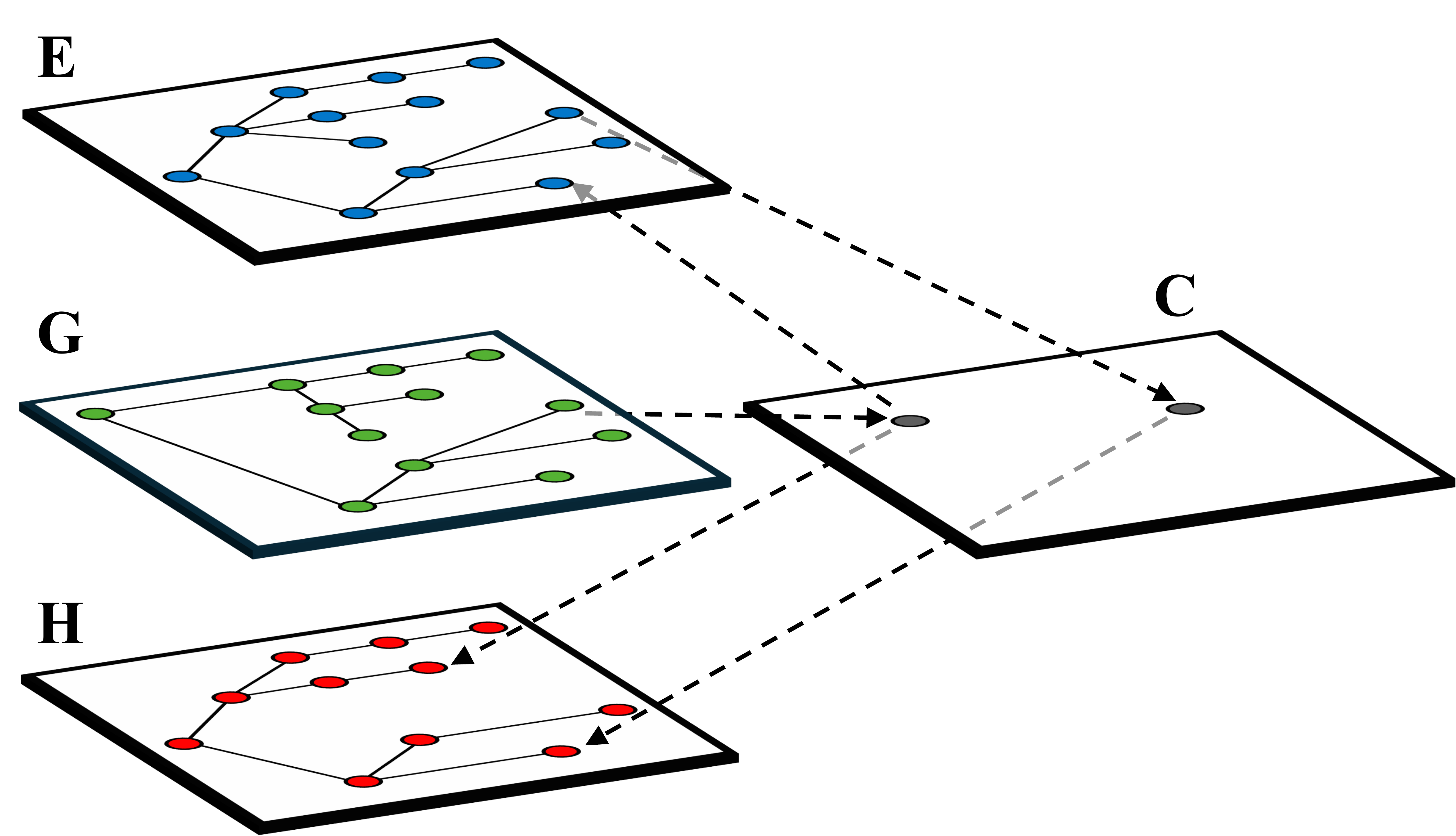}
    \caption{Graphical Representation of the Multi-Layer MEN Structure.}
    \label{fig:MEN}
\end{figure}

The proposed framework, shown in Fig.~\ref{fig:MEN}, models electricity, gas, and heat networks as individual layers in a unified multi-layer graph. Each layer contains domain-specific \textit{nodes} (\textit{e.g.}, electrical buses, gas junctions, thermal substations) and \textit{edges} (\textit{e.g.}, transmission lines, pipelines, heat exchangers.

To enable interaction between energy carriers, a dedicated coupling layer is introduced. This layer contains inter-layer edges representing energy conversion technologies such as CHP units, P2G, Electrolyzer-to-Gas (E2G) systems, gas boilers, and electric heat pumps. Each of these devices is modeled as a directional connector between nodes of different domains (e.g., electric to heat), with technology-specific conversion efficiencies.

\begin{figure}[t]
    \centering
    \includegraphics[width=0.85\linewidth]{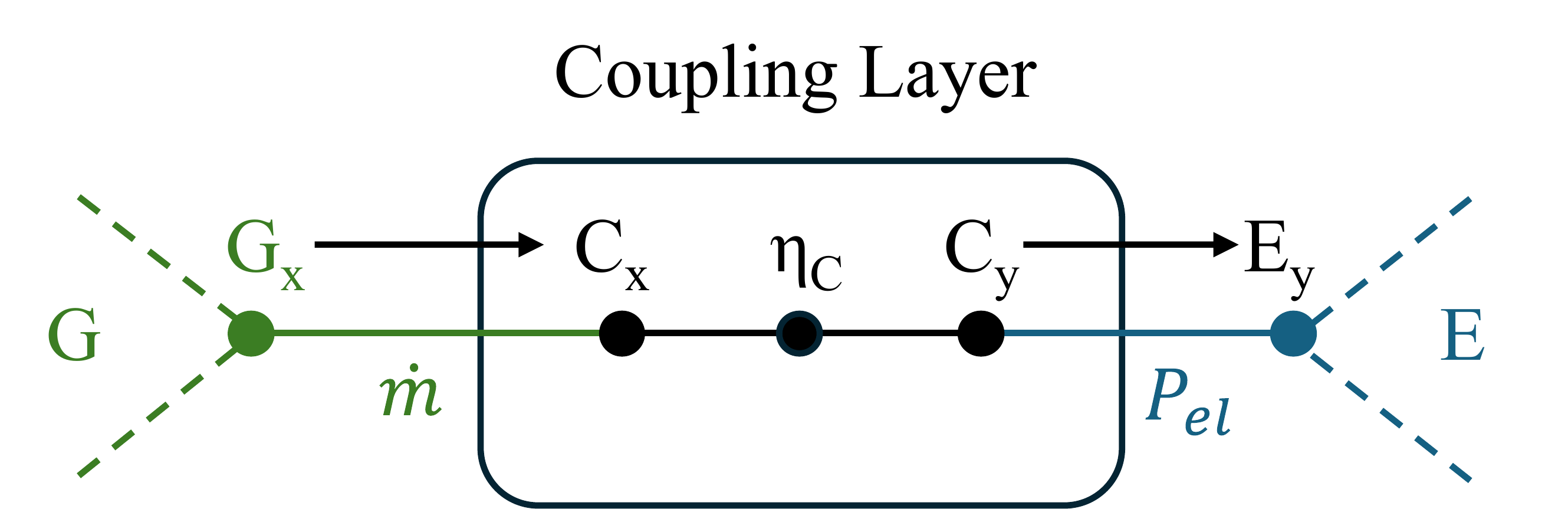}
    \caption{Schematic of Coupling Layer and Interdomain Node Mapping}
    \label{fig:coupling_schematic}
\end{figure}

A central modeling construct is the introduction of a dedicated coupling layer that enables interconnection between domains via node duplication. For each interconnection, the physical node (e.g., gas or electric bus) is replicated in the coupling layer and connected via a coupling edge. 
As shown in Fig.~\ref{fig:coupling_schematic}, the coupling network (C) mediates interactions between the gas (G) and electrical (E) networks through conversion devices, modeled by the following energy balance equations:
\begin{align}
\text{CHP:} \quad & Q_{\mathrm{th}} = \eta_{\mathrm{CHP}} \times \, P_{\mathrm{el}} \label{eq:CHP} \\
\text{E2G:} \quad & H_{\mathrm{gas}} = \eta_{\mathrm{E2G}} \times \, P_{\mathrm{el}} \label{eq:E2G} \\
\text{Gas boiler:} \quad & Q_{\mathrm{th}} = \eta_{\mathrm{GB}} \times \, H_{\mathrm{gas}} \label{eq:GB} \\
\text{Heat pump:} \quad & Q_{\mathrm{th}} = \mathrm{COP}(T) \times \, P_{\mathrm{el}} \label{eq:HP}
\end{align}


The coupling layer is fully integrated into the system graph and formulated analogously to the individual energy domains. Through explicit node duplication and edge-based conversion modeling, the approach maintains a consistent topological structure, while enabling accurate and scalable representation of sector-coupling processes. All domain-level attributes used in the network formulation are summarized in Table~\ref{tab:men_attributes}.

\begin{table}[b]
\centering
\caption{Key attributes in the modeled MEN}
\begin{tabular}{l l P{0.60\columnwidth}}
\toprule
\textbf{Domain} & \textbf{Entity} & \textbf{Selected Attributes} \\
\midrule
\multirow{2}{*}{\parbox{1.5cm}{\raggedright Electricity}}
  & Node & \texttt{voltage}, \texttt{vm\_pu}, \texttt{va\_degree}, \texttt{load} \\
  & Edge & \texttt{r\_ohm\_per\_km}, \texttt{length\_km} \\
\midrule
\multirow{2}{*}{\parbox{1.5cm}{\raggedright Gas}}
  & Node & \texttt{pn\_bar}, \texttt{mdot}, \texttt{tfluid\_k}, \texttt{height\_m} \\
  & Edge & \texttt{diameter\_m}, \texttt{length\_km}, \texttt{k\_mm}, \texttt{mdot} \\
\midrule
\multirow{2}{*}{\parbox{1.5cm}{\raggedright Heat}} 
  & Node & \texttt{pn\_bar}, \texttt{tfluid\_k}, \texttt{mdot}, \texttt{type} \\
  & Edge & \texttt{diameter\_m}, \texttt{length\_km}, \texttt{heat\_w} \\
\bottomrule
\end{tabular}
\label{tab:men_attributes}
\end{table}

\subsection{Multi-Energy Flow Calculation}
\label{Solution}

This section presents the formulation of the power flow problem for MESs, comprising electricity, gas, and heat systems. Each domain is modeled by steady-state equations, which are solved simultaneously using the NR method. The key innovation lies in assembling the domain-specific Jacobians into a unified block structure, enabling simultaneous solution and consistent convergence behavior across carriers.

The NR algorithm linearizes the nonlinear steady-state equations for each domain and solves them iteratively. Below, we detail the formulations per energy carrier.

\subsubsection{Electrical Power Flow}

The electrical subsystem follows classical power flow equations based on nodal admittance $\underline{Y}_{ik}$:
\begin{equation}
    \underline{I}_i = \sum_{k=1}^{N} \underline{Y}_{ik} \cdot \underline{V}_k,
    \label{eq: knotenstrom}
\end{equation}
\begin{equation}
    \underline{S}_i = \underline{V}_i \cdot \underline{I}_i^*,
    \label{eq: mat scheinleistung}
\end{equation}
where $\underline{I}_i$ is the current, $\underline{V}_k$ is the voltage and $\underline{S}_i$ is the apparent power. The system is linearized with the Jacobian matrix:
\begin{equation}
    \begin{bmatrix}
        J_{11}^\text{E} & J_{12}^\text{E} \\
        J_{21}^\text{E} & J_{22}^\text{E}
    \end{bmatrix}
    \cdot \Delta
    \begin{bmatrix}
        \mathbf{|V|}\\
        \boldsymbol{\delta}
    \end{bmatrix}
    =
    \begin{bmatrix}
        \Delta \mathbf{P} \\
        \Delta \mathbf{Q}
    \end{bmatrix}.
    \label{eq:jacobi_matrix}
\end{equation}
Here, $\mathbf{|V|}$ and $\boldsymbol{\delta}$ are the voltage magnitudes and phase angles at each node, while $\Delta \mathbf{P}$ and $\Delta \mathbf{Q}$ are the active and reactive power mismatches.

\subsubsection{Gas Flow}

Mass flow in the gas network satisfies both continuity and pressure drop constraints:
\begin{equation}
    \sum_{b = 1}^{N_b} \dot{m}_{b} + \dot{m}_n = 0,
    \label{eq: mdot knoten}
\end{equation}
\begin{equation}
    p_\text{in} - p_\text{out} = \frac{\rho v^2}{2} \left( \frac{\lambda l}{d} + \zeta \right) - \rho g \Delta h,
    \label{eq: druck bilanz}
\end{equation}
where $\dot{m}$ is the mass flow rate, $p$ is the nodal pressure, $\rho$ is the gas density, and $v$ is the flow velocity. The parameter $\lambda$ denotes the friction factor, $l$ is the pipe length, $d$ is the internal diameter, and $\zeta$ accounts for local loss coefficients due to valves or fittings. The term $\Delta h$ represents the elevation difference between nodes, and $g$ is gravitational acceleration.

These equations result in the gas-domain NR formulation:
\begin{equation}
    \begin{bmatrix}
        J_{11}^\text{G} & J_{12}^\text{G} \\
        J_{21}^\text{G} & J_{22}^\text{G}
    \end{bmatrix}
    \cdot \Delta
    \begin{bmatrix}
        \mathbf{p_n} \\
        \dot{\mathbf{m}}_b
    \end{bmatrix}
    =
    \begin{bmatrix}
        \Delta \dot{\mathbf{m}}_n \\
        \Delta \mathbf{p}_b
    \end{bmatrix},
    \label{eq: jacobimatrix gas}
\end{equation}
where $\mathbf{p_n}$ are the nodal pressures and $\dot{\mathbf{m}}_b$ are the edge-based mass flows.

\subsubsection{Thermal Flow}

The heating network is governed by hydraulic flow -- formulated analogously to the gas network \eqref{eq: jacobimatrix gas} -- and thermal energy propagation dynamics. Energy conservation at nodes and pipe segments is enforced through:

\begin{equation}
    \sum \dot{Q}_{\text{in}} - \sum \dot{Q}_{\text{out}} - \dot{Q}_{\text{loss}} = 0,
    \label{eq: energiebilanz heat}
\end{equation}
\begin{equation}
    \dot{m} \cdot c_p \cdot (T_{b{\text{,in}}} - T_{b{\text{,out}}}) - \dot{Q}_{\text{loss}} - \dot{Q}_{\text{heat}} = 0,
    \label{eq: leitungbilanz heat}
\end{equation}
where $\dot{m}$ is the mass flow rate, $c_p$ is the specific heat capacity of water, $T_{b,\text{in}}$ and $T_{b,\text{out}}$ are the inlet and outlet temperatures, $\dot{Q}_{\text{loss}}$ denotes heat loss to the environment, and $\dot{Q}_{\text{heat}}$ is the thermal power delivered to or extracted from the system.

These equations yield the NR formulation for the thermal domain:
\begin{equation}
    \begin{bmatrix}
        J_{11}^{\text{H}_{\text{th}}} & J_{12}^{\text{H}_{\text{th}}} \\
        J_{21}^{\text{H}_{\text{th}}} & J_{22}^{\text{H}_{\text{th}}}
    \end{bmatrix}
    \cdot \Delta
    \begin{bmatrix}
        \mathbf{T_n} \\
        \mathbf{T_{b\text{,out}}}
    \end{bmatrix}
    =
    \begin{bmatrix}
        \Delta \dot{\mathbf{Q}}_n \\
        \Delta \dot{\mathbf{Q}}_b
    \end{bmatrix},
    \label{eq: jacobimatrix thermal}
\end{equation}
where $\mathbf{T_n}$ are the nodal temperatures, $\mathbf{T_{b,\text{out}}}$ are the outlet temperatures of each branch.

\subsubsection{Coupled Multi-Energy System}

The full MES is solved using a block Jacobian matrix that integrates all domain-specific NR equations. The resulting structure enables simultaneous solution of electricity, gas, and heat variables:

\setlength{\arraycolsep}{0.1pt} 
\begin{equation}
\left[
\begin{array}{cc cc cc cc}
J_{11}^{\text{E}} & J_{12}^{\text{E}} & 0 & 0 & 0 & 0 & 0 & 0\\
J_{21}^{\text{E}} & J_{22}^{\text{E}} & 0 & 0 & 0 & 0 & 0 & 0\\
0 & 0 & J_{11}^{\text{G}} & J_{12}^{\text{G}} & 0 & 0 & 0 & 0\\
0 & 0 & J_{21}^{\text{G}} & J_{22}^{\text{G}} & 0 & 0 & 0 & 0 \\
0 & 0 & 0 & 0 & J_{11}^{\text{H}_{\text{hy}}} & J_{12}^{\text{H}_{\text{hy}}} & 0 & 0\\
0 & 0 & 0 & 0 & J_{21}^{\text{H}_{\text{hy}}} & J_{22}^{\text{H}_{\text{hy}}} & 0 & 0\\
0 & 0 & 0 & 0 & 0 & 0 &J_{11}^{\text{H}_{\text{th}}} & J_{12}^{\text{H}_{\text{th}}}\\
0 & 0 & 0 & 0 & 0 & 0 &J_{21}^{\text{H}_{\text{th}}} & J_{22}^{\text{H}_{\text{th}}}\\
\end{array}
\right]
\cdot
\left[
\begin{array}{c}
\mathbf{|V|} \\
\boldsymbol{\delta} \\
\mathbf{p_n} \\
\dot{\mathbf{m}}_b \\
\mathbf{p}_n^{\text{H}} \\
\dot{\mathbf{m}}_b^{\text{H}} \\
\mathbf{T}_n \\
\mathbf{T}_{b\text{,out}}
\end{array}
\right]
\mathbin{=}
\left[
\begin{array}{c}
\Delta \mathbf{P} \\
\Delta \mathbf{Q} \\
\Delta \dot{\mathbf{m}}_n \\
\Delta \mathbf{p}_b \\
\Delta \dot{\mathbf{m}}_n^{\text{H}} \\
\Delta \mathbf{p}_b^{\text{H}} \\
\Delta \dot{\mathbf{Q}}_n \\
\Delta \dot{\mathbf{Q}}_b
\end{array}
\right]
\label{eq:jacobi_blocksystem_fixed}
\end{equation}

In this formulation, each Jacobian submatrices contain the partial derivatives of a specific domain's residual equations with respect to its own state variables. These derivatives are updated at each iteration and updated to reflect the system's nonlinear behavior. The block structure captures intra-domain interactions while the coupling layer models inter-domain dependencies, enabling a unified and integrated NR solution approach.

\section{Case Study}
\label{sec:case}

\begin{figure}[t]
    \centering
    \includegraphics[width=\linewidth]{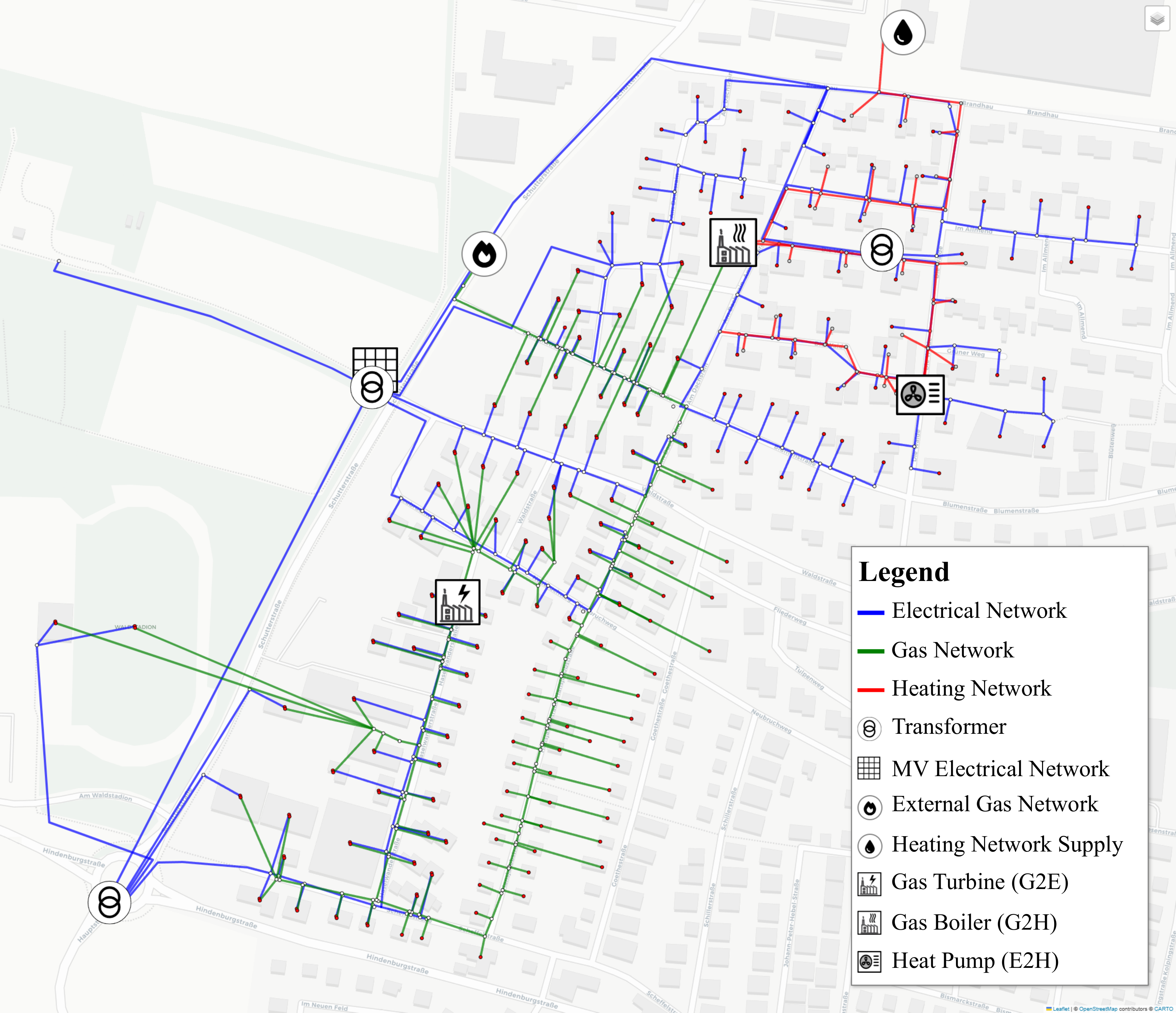}
    \caption{Geographical layout of the MEN including coupling elements on OpenStreetMaps background}
    \label{fig:netz koppel full}
\end{figure}

To demonstrate the modeling and simulation of a MES, a realistic example based on data from the Schutterwald region \cite{kisseGISBasedPlanningApproach2020} is constructed. It integrates representative electricity, gas, and district heating networks, coupled through gas generators, gas boilers, and heat pumps. The overall topology is visualized in Fig.~\ref{fig:netz koppel full}.

Table~\ref{tab:case strom anzahl} summarizes the number of nodes, edges, and loads in each subnetwork. The subnetworks are interconnected through interface components.

\begin{table}[b]
    \centering
    \caption{Overview of elements in the MEN.}
    \begin{tabular}{l c c c c }
        \toprule
        \textbf{Element} & \textbf{Electricity} & \textbf{Gas} & \textbf{Heat} \\
        \midrule
        Nodes & 295 & 236 & 132 \\
        Edges & 291 & 235 & 106 \\
        Loads & 187 & 123 & 25 \\
        \bottomrule
    \end{tabular}
    \label{tab:case strom anzahl}
\end{table}

\subsection{Electrical Network.}  
The electrical network is based on a modified low-voltage \SI{0.4}{kV} and medium-voltage \SI{20}{kV} distribution network derived from Schutterwald. Standard cable types used in the distribution network include \texttt{NAYY 4$\times$50 SE} and \texttt{NAYY 4$\times$150 SE} for low-voltage (LV) segments, and \texttt{NA2XS2Y 1$\times$185 RM/25} for medium-voltage (MV) segments. The corresponding impedance parameters are adopted from representative manufacturer datasheets.
Three transformers connect the voltage levels, with rated capacities ranging from 0.25 MVA to 0.63 MVA and a short-circuit voltage of \( v_{sc} = 6\% \). Capacitive effects are neglected for the sake of simplification. The slack bus voltage magnitude is fixed at \( |V| = 1 \) p.u., with a phase angle of \( \delta = 0^\circ \).

\subsection{Gas Network.}  
The gas network is modeled using standard \texttt{PE 100 SDR 11} pipelines with diameters ranging from \SI{50}{\milli\meter} (house connections) to $147\,\si{\milli\meter}$ and an assumed wall roughness of $0.1\,\si{\milli\meter}$. The model is based on geospatial and topographical data obtained from STANET, incorporating detailed elevation profiles. A steady-state, isothermal flow assumption with $T_\text{gas} = \SI{10}{\degreeCelsius}$ is applied, which is valid due to low flow velocities and typical burial depths. A fixed-pressure reference node supplies the network, and a representative subnet is extracted from the full model for analysis.

\subsection{District Heating Network.}
The heat network utilizes \texttt{PMR-type} pipes and is driven by a central circulation pump providing a steady mass flow rate of \SI{9}{\kilogram\per\second} at a supply temperature of \SI{80}{\degreeCelsius}.

\subsection{Electric Loads.}  
Standard load profiles from \texttt{SimBench 2016} \cite{meineckeSimBenchBenchmarkDataset2020} are assigned to residential and commercial buildings to capture diverse consumption behaviors such as weekday operations, evening peaks, or weekend loads. A fixed power factor of $\cos(\phi) = 0.93$ (inductive) is assumed uniformly  for all electrical loads.

\subsection{Thermal Loads.}  
Heat demand profiles are generated using the \texttt{BDEW method} implemented through the \texttt{demandlib} package \cite{oemofDeveloperGroup.2016}. The methodology differentiates between single- and multi-family residences as well as various commercial buildings. Typical peak heat loads range from $6.8\,\si{\kilo\watt}$ (modern homes) to $200\,\si{\kilo\watt}$ (workshops), based on standardized heat density and floor area assumptions.

The MEN is completed by three interface technologies:
\begin{itemize}
    \item {Gas Turbine (\texttt{G2E})}: \( P_{\mathrm{el}} = \SI{60}{kW}, \quad \eta = 0.5 \)
    \item {Gas Boiler (\texttt{G2H})}: \( \dot{Q}_{\mathrm{th}} = \SI{73}{kW}, \quad \eta = 0.977 \)
    \item {Heat Pump (\texttt{E2H})}:  \( \dot{Q}_{\mathrm{th}} = \SI{76.7}{kW}\), with a dynamic COP based on meteorological data
\end{itemize}
Each unit operates using normalized time series derived from aggregated building demand.

\section{Results and Discussion}
\label{Results}

The developed coupled solver was tested on a representative MES under two seasonal scenarios. For brevity, we report results for the coldest week of the year, during which peak thermal loads coincided with low ambient temperatures. A worst-case operational snapshot was identified at 08:00 on Friday, January 22.

\begin{figure}[t]
    \centering
    \includegraphics[width=0.85\linewidth]{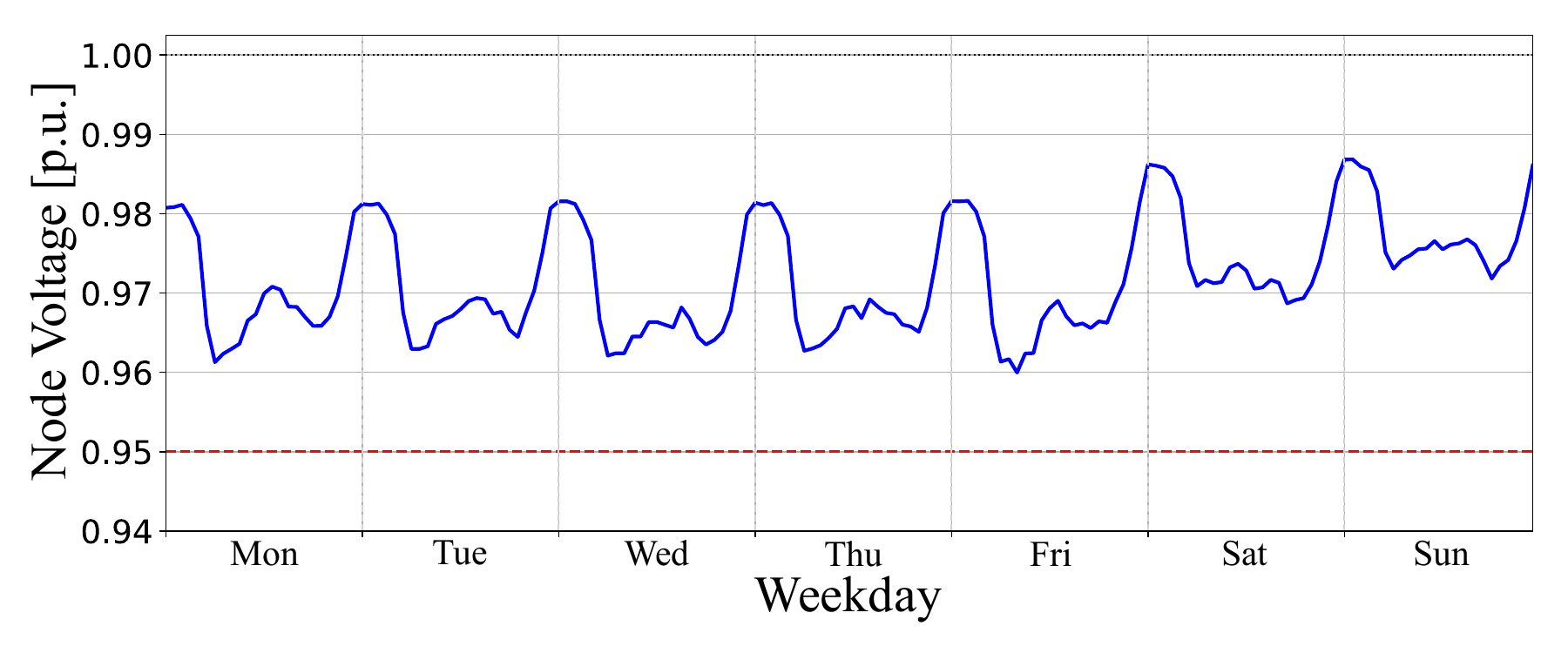}
    \caption{Minimum node voltages for a winter week (Jan 18–24).}
    \label{fig: box knoten strom}
    \centering
    \includegraphics[width=0.85\linewidth]{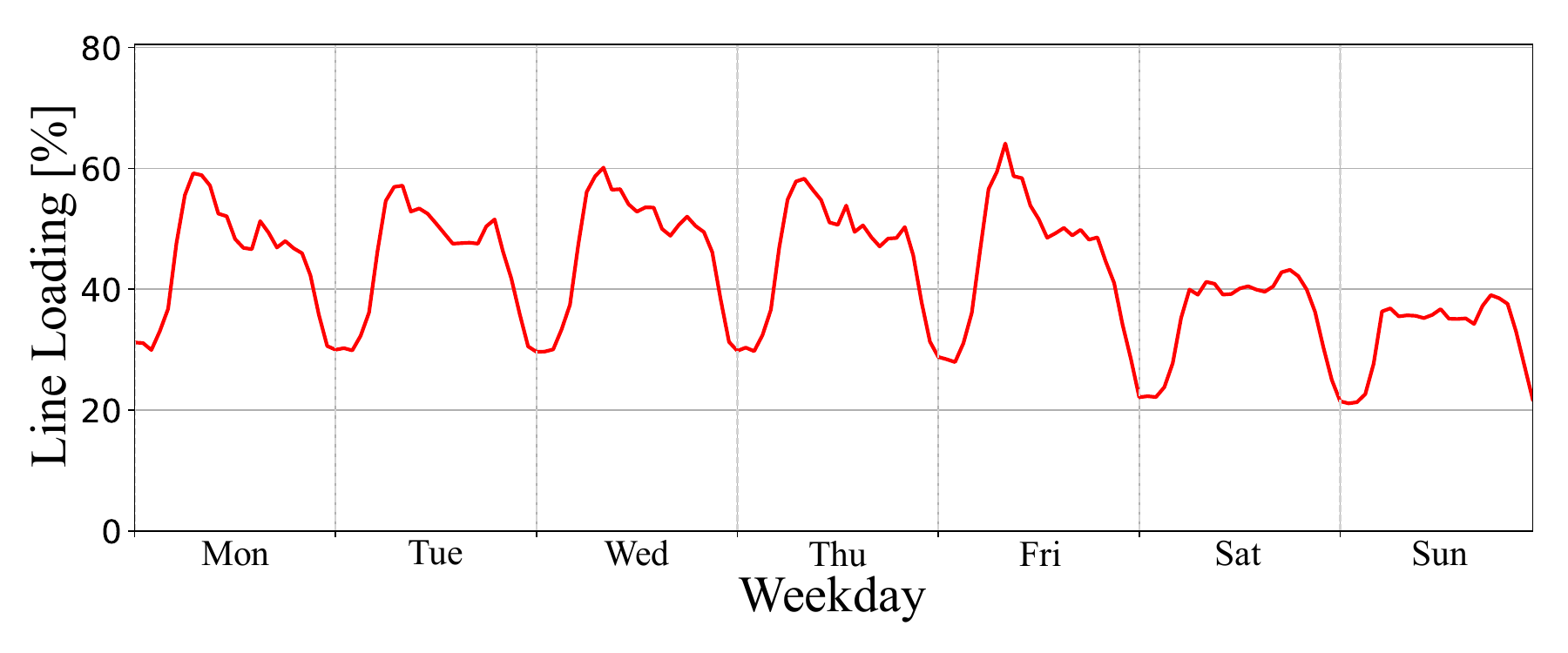}
    \caption{Maximum line loading for a winter week (Jan 18–24).}
    \label{fig: box kanten strom}
\end{figure}

The results confirm the correct interaction of the coupled domains and the ability of the solver to maintain steady-state convergence throughout the simulated period. In the electrical domain, all nodal voltages remained above \SI{0.96}{\perunit} (Figure~\ref{fig: box knoten strom}), and transformer and line loadings stayed below $70\,\%$ (Figure~\ref{fig: box kanten strom}). Pressure in the gas network dropped by less than $1\,\%$ from the reference node, with peak mass flows occurring during cold morning hours. The district heating network maintained forward supply temperatures within the operational range of $\SIrange{75}{90}{\degreeCelsius}$, while return lines showed slight overheating due to combined feed-in from gas boilers and heat pumps.

\begin{figure*}[ht]
    \centering
    \begin{subfigure}[b]{0.32\textwidth}
        \includegraphics[width=\textwidth]{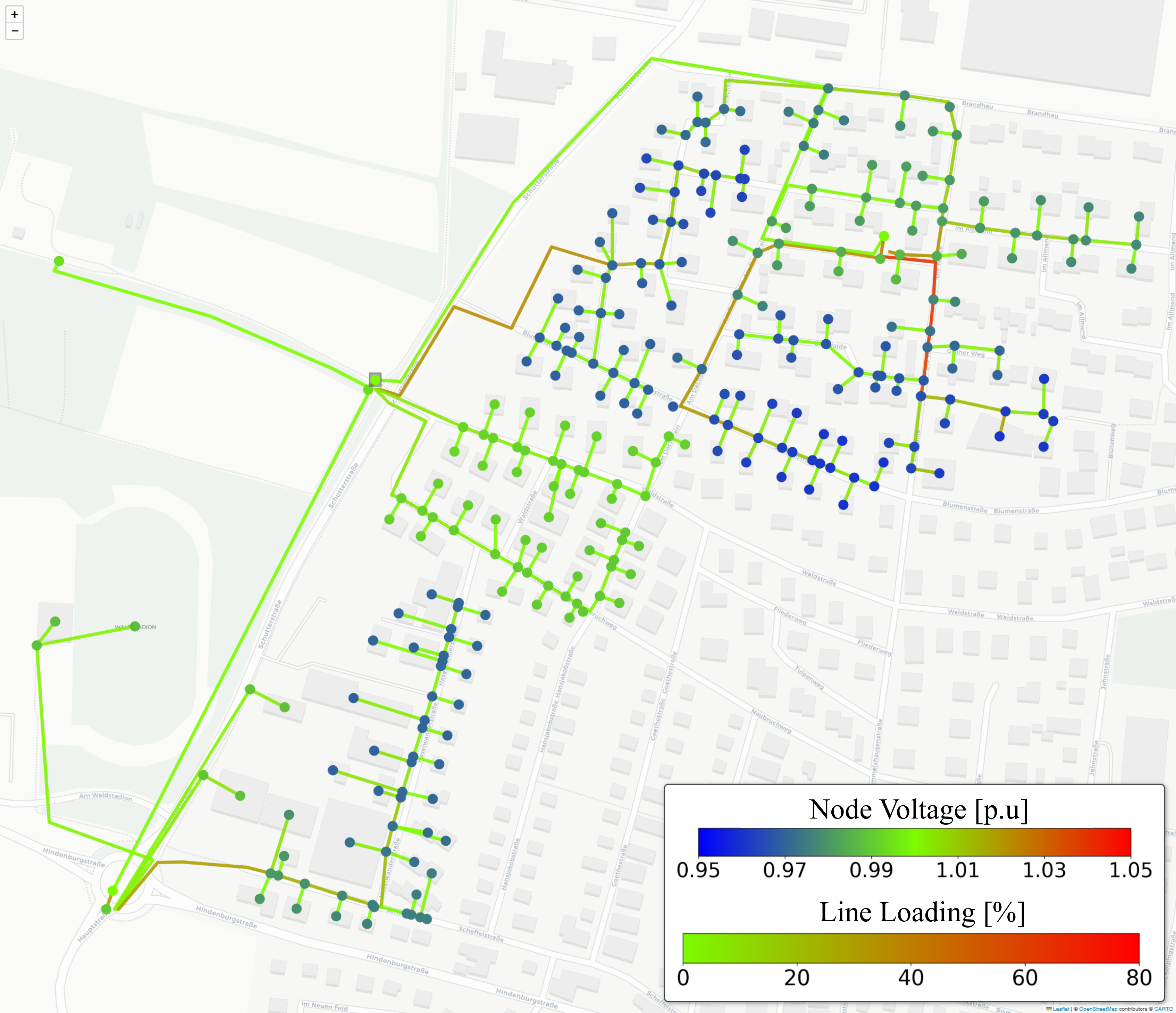}
        \caption{Electrical network results}
        \label{fig:elec_heatmap}
    \end{subfigure}
    \hfill
    \begin{subfigure}[b]{0.32\textwidth}
        \includegraphics[width=\textwidth]{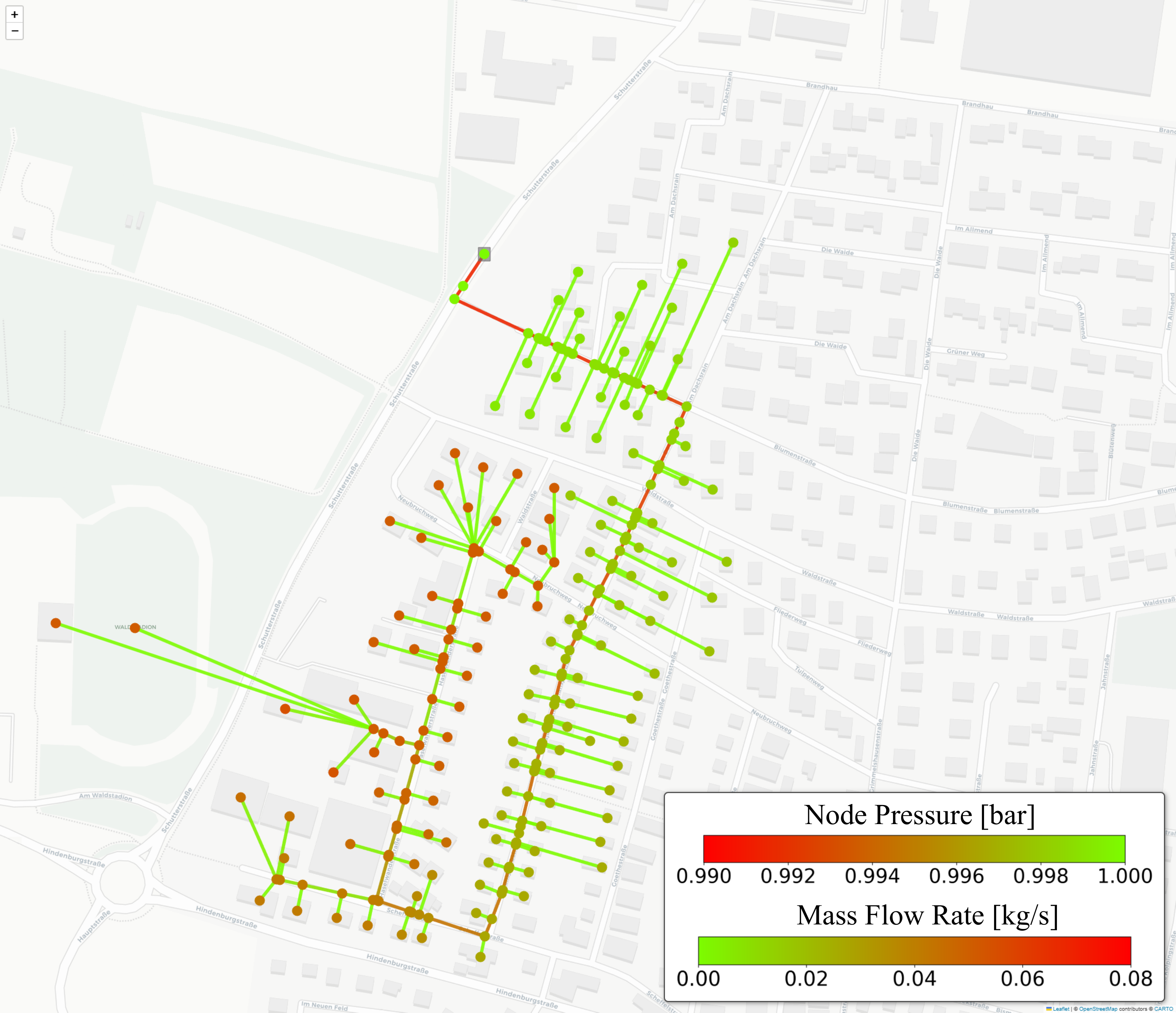}
        \caption{Gas network results}
        \label{fig:gas_heatmap}
    \end{subfigure}
    \hfill
    \begin{subfigure}[b]{0.32\textwidth}
        \includegraphics[width=\textwidth]{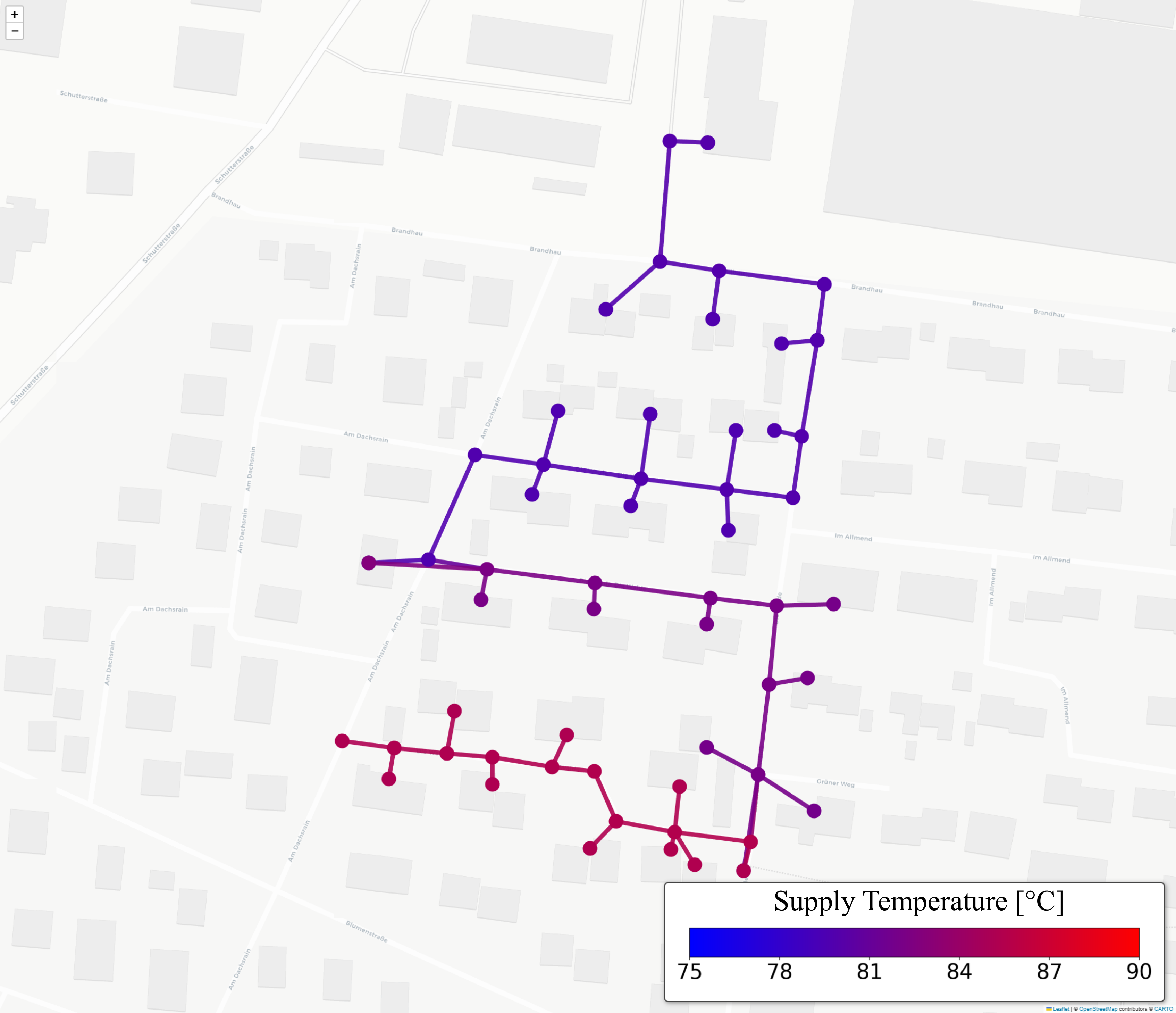}
        \caption{District heating network results}
        \label{fig:heat_heatmap}
    \end{subfigure}
    \caption{Power Flow Simulation Results for Peak Hour on Jan 22, 08:00 across Energy Domains (a–c).}
    \label{fig:result_heatmap}
\end{figure*}

These results are spatially illustrated in Figure~\ref{fig:elec_heatmap} for the electrical network, Figure~\ref{fig:gas_heatmap} for the gas network, and Figure~\ref{fig:heat_heatmap} for the district heating network for the time step with peak loading at 08:00 on Friday, January 22.

\subsection{Computational Performance}
Solver accuracy was benchmarked against established tools (\texttt{pandapower} + \texttt{pandapipes}), showing negligible deviations (\textit{e.g.}, voltage error $<10^{-8}$\,p.u.). Table~\ref{tab:comparison} summarizes runtime and numerical accuracy. While slower than decoupled solvers, the coupled method offers a consistent solution across all domains in a unified formulation. Results show that approximately 98\% of the total computation time is spent on data preprocessing and interfacing, while less than 2\% is used for the numerical solution. This shows a substantial runtime improvement potential through improved data management.

\begin{table}[t]
\centering
\caption{Performance of the proposed method vs. reference tools.}
\label{tab:comparison}
\begin{tabular}{l c c}
\toprule
\textbf{Metric} & \textbf{Proposed Method} & \textbf{pandapower+pandapipes} \\
\midrule
Avg. solving time  & $133.4$ [ms] & $39.6$* [ms] \\
Avg. iterations no. & $2.9$ & --- \\
\midrule
Voltage error & $9.1 \times 10^{-9}$ & reference \\
Pressure error       & $1.5 \times 10^{-11}$ & reference \\
Temperature error  & $3.6 \times 10^{-6}$ & reference \\
\bottomrule
\end{tabular}
\begin{flushleft}\footnotesize
*Reference runtime is the sum of individual electricity ($19.1$ ms), gas ($7.1$ ms) and heat ($13.3$ ms) solves executed sequentially. 
\end{flushleft}
\end{table}

\section{Conclusion and Outlook}
\label{Conclusion}

This paper presented a unified modeling and solution approach for steady-state analysis of MES, integrating electricity, gas, and heat domains into a consistent mathematical framework. Each domain was modeled as a graph, and interconnections were captured through a coupling layer, enabling simultaneous solution of the entire system using the NR method.
The proposed method was validated through a case study based on real-world data. The solver demonstrated convergence across all time steps and produced results consistent with established domain-specific tools, confirming both the validation of the formulation and its numerical performance. The coupled structure allows for the representation of interactions between energy carriers and enables holistic system assessments within a single calculation step.

Future work will focus on extending the framework toward optimization, enabling sector coupling-aware network expansion and reinforcement planning.


\balance

\bibliographystyle{IEEEtran}
\bibliography{ISGT25.bib}

\end{document}